\documentstyle[emulateapj,psfig,apjfonts]{article}

\def\gs{\mathrel{\raise0.35ex\hbox{$\scriptstyle >$}\kern-0.6em 
\lower0.40ex\hbox{{$\scriptstyle \sim$}}}}
\def\ls{\mathrel{\raise0.35ex\hbox{$\scriptstyle <$}\kern-0.6em 
\lower0.40ex\hbox{{$\scriptstyle \sim$}}}}

\begin{document}

\submitted{ApJ in press}

\title{A low global star formation rate in the rich galaxy cluster 
AC\,114 at $z=0.32$}

\author{Warrick J.\ Couch$^1$, Michael L. Balogh$^2$, Richard
G. Bower$^2$, Ian Smail$^2$,\\ Karl Glazebrook$^{3,4}$, Melinda Taylor$^1$} 
\affil{\small 1) School of Physics, University of New South Wales, Sydney
2052, Australia}
\affil{\small 2) Department of Physics, University of Durham, South Road,
Durham, DH1 3LE, UK}
\affil{\small 3) Anglo-Australian Observatory, P.O. Box 296, Epping, NSW
2121, Australia}
\affil{\small 4) Department of Astrophysics, Johns Hopkins University,
Baltimore, Maryland, USA}

\begin{abstract}
We present the results of a wide-field survey for H$\alpha$-emitting
galaxies in the  cluster AC\,114 at $z=0.32$. Spectra centred on
H$\alpha$ at the cluster redshift have been obtained for 586 galaxies
to $I_{tot}\sim 22$ out to a radius of $\sim 2\,h_{50}^{-1}$\,Mpc. 
At most, only $\sim 10$\% of these were found to be H$\alpha$--emitting cluster
members. These objects are predominantly blue and of 
late-type spiral morphology, consistent with them hosting star 
formation. However, $\sim 65$\% of the cluster members classified
morphologically as spirals (with {\it HST}), have no detectable H$\alpha$
emission; star-formation and morphological evolution in cluster
galaxies appear to be largely decoupled. Changes in the H$\alpha$
detection rate and the strength of H$\alpha$ emission with environment
(as traced by local galaxy density) are found to be weak
within the region studied. Star formation within the cluster members
is also found to be strongly {\it and} uniformly suppressed with the
rates inferred from the H$\alpha$ emission not exceeding 
4\,M$_{\odot}$\,yr$^{-1}$, and AC\,114's H$\alpha$ luminosity function
being an order of magnitude below that observed for field galaxies at the
same redshift. None of the galaxies detected have the high star formation
rates associated with `starburst' galaxies; however, this may still
be reconciliable with the known ($8\pm 3$\%) fraction of `post-starburst' 
galaxies within AC\,114, given the poorly determined but short lifetimes
of starbursts and the possibility that much of the associated star
formation is obscured by dust. 
\end{abstract}


\section{Introduction}

Clusters of galaxies provide a powerful laboratory for studying the
evolution of galaxies. In the hierarchical picture of galaxy formation,
clusters grow in mass by accreting galaxies from the surrounding field.
Once these galaxies enter the dense environment of the cluster, star
formation is suppressed and the galaxies evolve to become early-type
galaxies.   Many processes operate to make the cluster environment
hostile to star forming galaxies. The most important processes
include:  mergers and violent encounters between galaxies (Lavery \&
Henry, 1994); tidal stripping and `harassment' (Byrd \& Valtonen 1990,
Moore et al.\ 1996); ram pressure stripping of the gas disk (Gunn \&
Gott 1972, Abadi et al.\ 1999, Quilis et al.\ 2000); and simply the
removal of any gas reservoir surrounding each galaxy (thus preventing
replenishment of the gas disk; Larson, Tinsley \& Caldwell 1980, Balogh
et al.\ 1999).

Since clusters can be observed out to $z>1$, we can use the look-back
times of the distant systems to directly investigate the evolution of
cluster galaxies. Butcher \& Oemler (1978a) and many subsequent authors
have noted that the fraction of blue galaxies in cluster cores
increases with redshift. Spectroscopic surveys have confirmed that
these blue galaxies are indeed cluster members (e.g.\ Dressler \& Gunn
1982, 1983; Couch \& Sharples 1987, hereafter CS; Lavery \& Henry 1988;
Dressler et al.\ 1999) and have identified a conspicuous population of
galaxies with no detectable emission (e.g.\ [O{\sc ii}]$\lambda$3727)
but with abnormally strong Balmer absorption lines (termed `E+A'
galaxies by Dressler \& Gunn, 1982 and `k+a' or `a+k' galaxies by
Dressler et al.\ 1999). It has also been proposed that the
systematically higher blue galaxy fractions in distant clusters is
coupled with a reduction in the S0 galaxy populations in these systems
[identified from imaging with the {\it Hubble Space Telescope} ({\it
HST})] in comparison to their local counterparts (Dressler et
al.\ 1997; Couch et al.\ 1998, hereafter C98).

The evolution of the galaxy populations in distant clusters may be
explained as the result of the increased activity of field galaxies at
intermediate redshift (Lilly et al.\ 1996), combined with a higher
accretion rate in these more distant systems (Bower 1991, Kauffmann
1996, Diaferio et al.\ 2000): not only are the galaxies forming stars
more rapidly when they arrive in the cluster, but they also arrive at a
higher rate.  However, while an overall picture of the evolution of
cluster galaxies is emerging, we have no detailed understanding of the
physical mechanism responsible for the suppression of star formation
and its effect on other properties such as mass and morphology.  The
key issues can be summarised by three questions:  (1)\,By how much and
how quickly is star formation suppressed in the infalling field
galaxies? (2)\,What is the radial dependence of the star formation
suppression? (3)\,Does the cluster simply suppress star formation or
does it first promote a burst of star formation?

The ability of current observations to address these issues is,
however, seriously hindered by most distant cluster studies having
largely concentrated on the central {\it core}
($r<1\,h_{50}^{-1}$\,Mpc) regions in which the BO-effect was originally
discovered: spectroscopic data of the quantity and quality needed to
track in detail the star formation characteristics of galaxies well
outside the core remain in short supply. Progress towards redressing
this problem has been made by Balogh et al.\ (1997, 1998), who used the
CNOC Survey data (Yee, Ellingson \& Carlberg 1996) to trace the
spectral properties of galaxies out to and beyond the virial radius in
15 X-ray bright clusters at $0.18<z<0.55$. Using the equivalent width
of the [O{\sc ii}]$\lambda$3727 emission line as a star formation
indicator, they found a smooth decline in the star formation rate from
the field into the interior of their clusters, an effect which they
claimed was likely to be independent of the morphology--density
relationship established by Dressler (1980). That this decline should
extend so far out in their clusters is surprising, since simple cluster
mass models would suggest that galaxies at and beyond the virial radius
should have only become recently bound to the cluster.

While the CNOC study is an important first step to understanding the
star formation patterns at large radii in distant clusters, a number of
issues remain. In particular, Poggianti et al.\ (1999; hereafter P99)
and Smail et al.\ (1999) have questioned the reliability of [O{\sc
ii}]$\lambda$3727 as a star formation indicator. P99 have noted that
there are insufficient bright, star-forming galaxies in their clusters
to explain the abundance of a+k/k+a objects. Instead, they suggest that
the progenitors of the a+k/k+a galaxies are e(a) emission line objects
(which show both weak [O{\sc ii}] emission and strong $H\delta$
absorption) in which the bulk of the star formation is obscured by dust
(see also Poggianti \& Wu 2000).  Furthermore, Smail et al.\ (1999)
suggest,  on the basis of radio continuum observations, that a
significant number of the apparently post-starburst (a+k/k+a) galaxies
in the core of Abell~851 ($z=0.41$) are in fact on-going star burst
galaxies in which the star formation (as measured by the [O{\sc ii}]
emission line) is completely hidden by dust.

Motivated by these issues and the clear need to explore further the
star formation behaviour at large cluster-centric radii, we have
embarked on an intensively--sampled, wide--field ($\sim
4\,h_{50}^{-1}$\,Mpc) H$\alpha$ survey of galaxies in 3 {\it
optically--selected} clusters at $z\sim 0.3$. These clusters -- AC103,
AC118, and AC\,114 -- have been subjected to extensive imaging and
spectroscopy both from the ground and from space ({\it HST} and {\it
ROSAT}), the results of which have shown them to be quite diverse in
their structural and X-ray properties. They therefore not only provide
useful targets for studying the cluster-to-cluster variations in star
forming galaxies, but also provide an important test of how the global
star formation history is dependent on cluster morphology. By
exploiting recent innovations in telescope instrumentation, our program
is able to survey large numbers of ($\sim 10^3$) galaxies in each
cluster field for H$\alpha$ emission -- the most direct indicator of
star formation -- down to very low, sub--1\,M$_{\odot}$\,yr$^{-1}$
levels. Hence it provides a highly complete inventory of star-forming
galaxies with sufficient statistical sampling at large cluster-centric
radii to robustly determine cluster-to-cluster trends.

In this paper we present the first observations to be made as part of this
program: H$\alpha$ spectroscopy of $\sim 600$ galaxies in the field of
AC\,114 at $z=0.32$. In the next section we describe our observations and
outline the novel techniques that allow us to obtain spectra for this
number of galaxies in just 4\,hrs on a 4-m telescope. In section 3, 
we provide details of our reduction techniques and the detection and
measurement of H$\alpha$ emission in our spectra. The numbers of H$\alpha$
emitting galaxies found, their distribution within the cluster, and the
star formation rates and luminosity function associated with this
population are then examined in section 4. We then discuss our results
in section 5 and summarise our conclusions in section 6.
 
\section{Observations}

The observations were made on the nights of 1999 August 14 and 15, using
the upgraded Low Dispersion Survey Spectrograph (LDSS$++$) on the 3.9\,m
Anglo-Australian Telescope. LDSS$++$ is a combined imager/spectrograph
whose performance has recently been enhanced by the installation of a new
red-optimised volume phase holographic (VPH) grism and the use of a
MIT/Lincoln Lab `deep depletion' $2048\times 4096$ pixel format CCD as
its detector. In combination, they give LDSS$++$ a factor of 2 better
throughput (at 7000\AA) than its predecessor when working in the `high
dispersion' (177\AA\,mm$^{-1}$, 9.5\AA\ resolution) mode used here (for
further details see Glazebrook 1998). 

In addition to these improvements, a `nod-and-shuffle' technique
has been implemented with LDSS$++$, providing for a significant increase
in multiplex gain when using the instrument in multi-slit mode. This
technique involves simultaneously shuffling the charge on the CCD  
in concert with moving the telescope back and forth on the sky (nodding),
so that the objects and sky are observed through the same part of the slit
and with the same pixels of the detector. It provides for high-precision
sky-subtraction -- particularly in the 0.7--1.0$\,\mu$m region where the
rapid time-sampling of the varying night-sky emission is a distinct
advantage --  and obviates the need for long slits (in order to
get good sampling of the sky adjacent to the object); consequently, it is
possible to work with very small apertures (`micro-slits') in the focal
plane mask and thus observe many more objects simultaneously. For this
study, this multiplex gain advantage was extended further  
by the use of a blocking filter to restrict our spectral coverage to the
wavelength window in which the H$\alpha$ emission from cluster members
would be seen: $8350<\lambda <8750$. As a result of these innovations, we
were able to observe up to $\sim 700$ galaxies with a single mask over
the $8.7'\times 8.7'$ ($3.1\times 3.1\,h_{50}^{-2}\,$Mpc$^{2}$ at
$z=0.32$, for $q_{0}=0.5$) field-of-view of LDSS$++$. Furthermore, a 
particular advantage of observing clusters at a redshift of $z\sim 0.3$ 
is that H$\alpha$ is redshifted into a wavelength region relatively clear 
of night sky emission lines.

In order to relate the H$\alpha$ emission to the underlying galaxy mass, 
our selection of spectroscopic targets was based on an $I$-band limited
galaxy sample identified within the field of AC\,114. Here we used the 
automated image detection and photometry package, SExtractor (Bertin \& 
Arnouts 1996), to generate a galaxy catalogue from a deep, AAT prime-focus
$I$-band image taken (for another project) in seeing of 1.2$''$ (FWHM). 
In doing so, a detection threshold of 1-$\sigma$ above sky and a minimum 
object area size of 10 pixels was used.   
A zero-point was established for the photometry by comparing our `total'
SExtractor magnitudes with calibrated $I$-band photometry derived for the
central $3.3'\times 2.1'$ region of AC\,114 from a CCD image taken as part
of the lensing study of Smail et al.\ (1991). This yielded a zero-point
with a 1$\sigma$ uncertainty of $\pm 0.04$\,mags. The final list of 
potential targets then consisted of all galaxies with $I_{tot}\leq 22.0$, 
supplemented by objects going $\sim 0.5$\,mags fainter than this limit to 
provide some flexibility in the mask design process. 

The multi-slit mask design was done semi-automatically using the {\sc
design} package purposely written by two of us (K.G.B.\ \& R.G.B.) for
general LDSS$++$ use. This takes the list of targets and their
astrometric positions, and optimally assigns slits to them to maximise
the number of objects observed. Here, we adopted a circular
`micro-slit' geometry -- a round hole 1.5$''$ in diameter
(corresponding to 9\,$h_{50}^{-1}$\,kpc at $z=0.32$) -- thought to be
the size and shape which best accommodated the typical seeing
experienced at the AAT (median of $\sim 1.2''$) and the desire to keep
aperture affects to a minimum. In running {\sc design}, some manual
intervention was required to ensure that a reasonable number of
galaxies from the spectroscopic samples of CS, Couch et al.\ (1994),
and C98 were included for comparison purposes. In addition, `double'
slits were assigned to about 30\% of the objects, allowing them to be
observed in both the `on-object' and `off-object' positions and thus
avoiding, in these cases, any reduction in on-source integration time
due to beam-switching. This involved placing a second slit 5$''$ east
of the object, being the location to which the telescope was moved for
the `off-object' exposure.

\centerline{\psfig{file=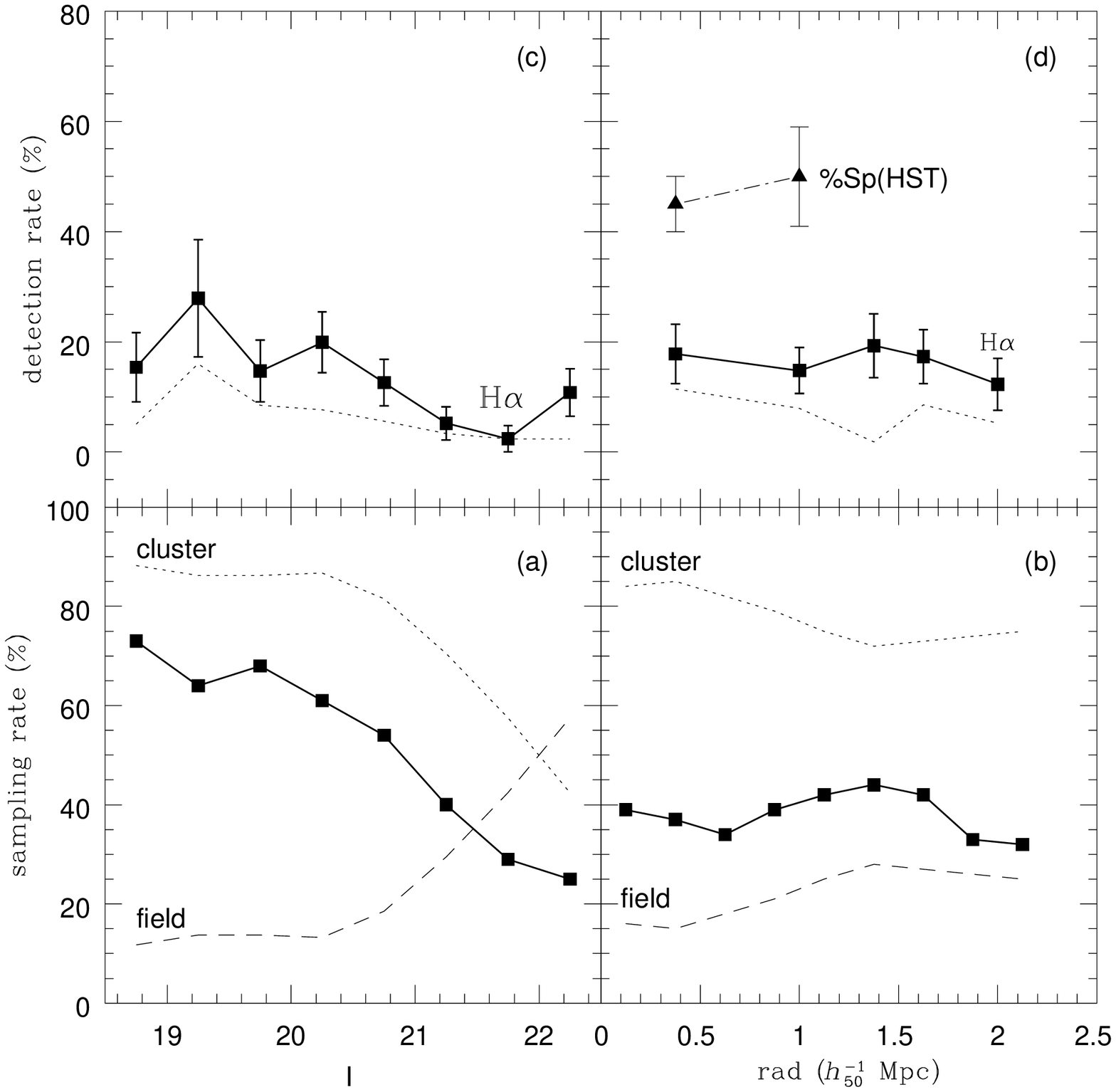,width=3.0in}}
\noindent{\small\addtolength{\baselineskip}{-1pt} 
{\sc Fig. 1.}-- Our sampling rates and H$\alpha$ detection rates as a
function of $I$ magnitude and radius from the centre of AC\,114. (a)The
connected {\it solid squares} indicate the percentage of galaxies
observed as a function of $I$ magnitude; the expected composition of
these galaxies in terms of being `cluster' and `field' objects is
indicated by the {\it dotted} and {\it dashed} lines, respectively.
(b)As for (a) but with clustercentric radius plotted on the abscissa.
(c)The detection rate of H$\alpha$-emitting `cluster' galaxies versus
$I$ magnitude. The connected {\it solid squares} indicate all
detections, be they case (i) or case (ii); the {\it dotted} line
indicates case (i) detections only. (d)As for (c) but as a function of
radius. The {\it dashed} line shows the rise with increasing radius in
the fraction of morphologically classified cluster spirals.

}

Upon completion of the mask design process, two masks were manufactured 
each containing slits for $\sim 630$ unique galaxies and $\sim 850$ slits
in total. Our photometric selection function for spectroscopic targets is
shown in Fig.~1(a) where it is seen that the sampling rate drops almost
monotonically from $\sim 70\%$ at $I_{tot}=19$ to 25\% at $I_{tot}=22$. In
terms of how our slits sample the galaxy distribution as a function of
radius from the centre of AC\,114, this is seen in Fig.~1(b) to be
reasonably constant at $\sim 40$\% out to $2\,h_{50}^{-1}$\,Mpc. For
comparison, we also show in Figs.~1(a) and 1(b) the expected composition
of our sample in terms of `cluster' and `field' galaxies as a function of
$I_{tot}$ and cluster-centric radius, respectively. These 
percentages (shown as the broken lines) are based on both number counts
from our $I$-band image and those of Metcalfe et al.\ (1995), the latter
being used to estimate the surface density of field galaxies. We see that
while the field contamination does rise with radius, it does so only 
slowly (going from $\sim 15$\% to $\sim 25$\%). This is due to AC\,114 being
significantly elongated in the SE-NW direction as can be seen in
Fig.~2. Here we show a combined $B+I$ grey-scale image of our LDSS$++$
field with contours showing the distribution of galaxies with $I\leq 20$
and which populate or lie close to AC\,114's E/S0 color-magnitude sequence
($1.7\leq B-I\leq 2.6$) overlayed. The contours show how flattened the
cluster is, with it extending right across the full diagonal of the
LDSS$++$ field. This flattened structure is also mirrored in the X-ray
emission from AC\,114 (Terlevich 1999).

Observations through only one of the two masks were possible, due to
the mixed weather conditions experienced on the 2 nights of our run. A
total of 4 hours integration, 2 hours `on-object' and 2 hours
`off-object', were obtained in clear conditions and seeing of
1.2--1.5$''$ (FWHM). This was broken up into a series of 30\,min
exposures during which 30 nod-and-shuffle cycles were executed, each
cycle involving a 30\,s integration in the `on-object' position, a 2\,s
interval in which the telescope was `nodded' and the charge on the CCD
shuffled, followed by another 30\,s integration in the `off-object' (or
`sky') position, and then a further 2\,s delay in order to nod the
telescope and shuffle the charge on the CCD back to its original
position. The performance of LDSS$++$ is such that in these 2\,hours of
on-source integration, a 3$\sigma$ flux limit of $2.0\times
10^{-17}$\,erg\,s$^{-1}$\,cm$^{-2}$ per 9.5\AA\ resolution element can
be reached, allowing us to detect cluster members with an H$\alpha$
luminosity of $L$(H$\alpha$)$=2.8\times
10^{40}\,h_{50}^{-2}$\,erg\,s$^{-1}$ and a star-formation rate of $\sim
0.25$\,M$_{\odot}$\,yr$^{-1}$.

\section{Reductions and H$\alpha$ measurements}

\setcounter{figure}{1}
\begin{figure*}
\centerline{\psfig{file=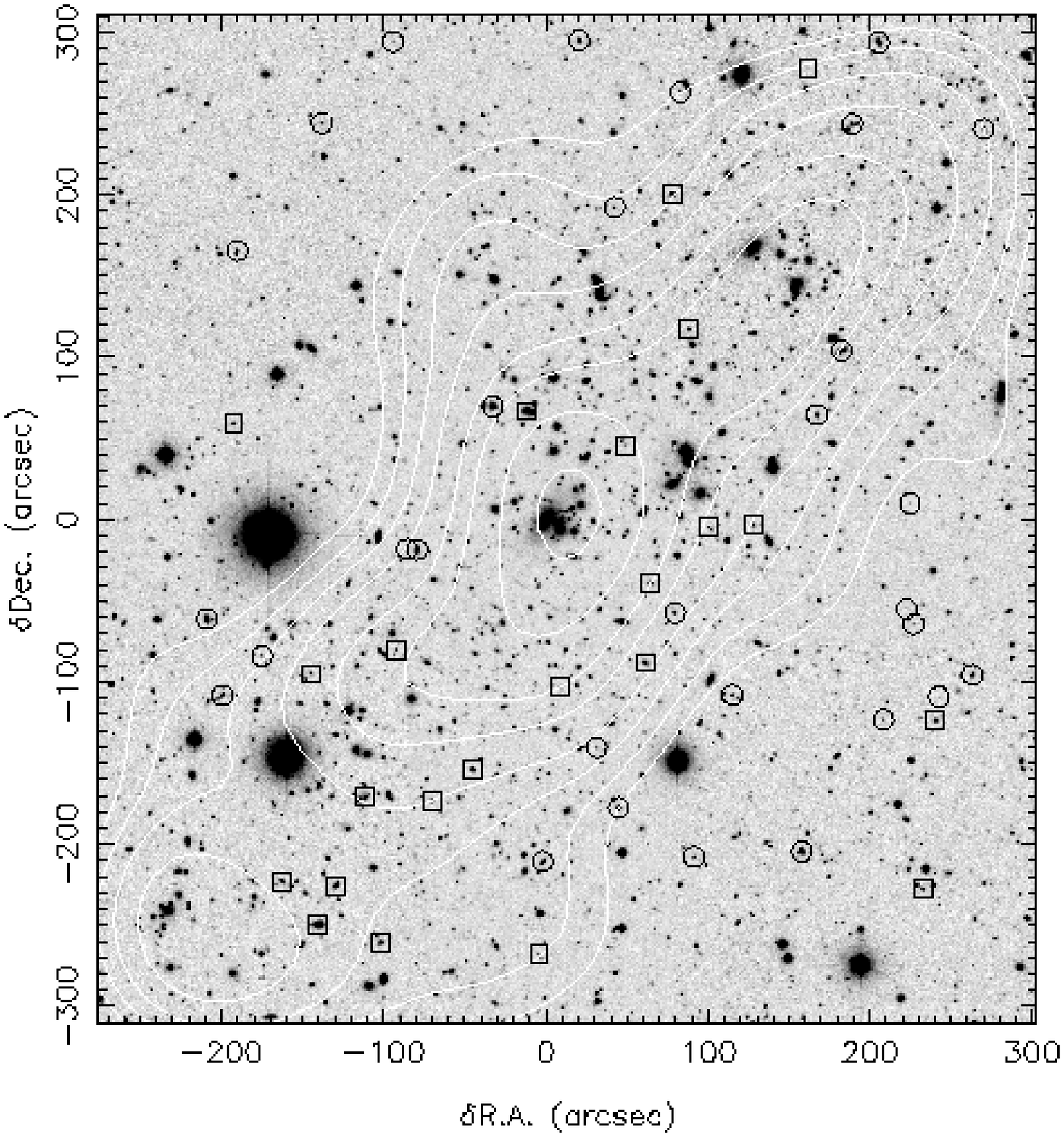}}
~\bigskip
\caption{A combined $B+I$ grey-scale image of the AC\,114 field observed 
with LDSS$++$. The {\it countours} represent the surface density 
distribution (smoothed using a Gaussian with $\sigma =
200\,h_{50}^{-1}$\,kpc) of galaxies within and close to AC\,114's E/S0 
color-magnitude sequence (see text for details), and span the interval
within 20\% of the maximum and minimum values in equal logarithmic steps.
The {\it squares} identify those galaxies with definite [case (i)]
H$\alpha$ detections while the {\it circles} identify those galaxies with
possible [case(ii)] H$\alpha$ detections. The size of the field is
$9.6'\times 10.2'$.} 
\end{figure*}

The final CCD image read out after each series of 30 nod-and-shuffle
cycles contains two equal $2048\times 1365$ pixel regions, one of which
has recorded the signal in the `on-object' position and the other the
signal in the `off-object' position. These were extracted separately
from all 8 of the CCD images obtained over our 4 hours of observation,
with the {\sc iraf} routine {\sc ficpix} used to interpolate over the
column defects introduced by the combination of trapping sites and the
charge shuffling process. They were then combined to produce two final
co-added `on-object' and `off-object' frames. The co-addition process
was performed using the {\sc imalign} and {\sc imcombine} routines
within {\sc iraf}; these ensured that there was perfect registration
between the individual frames and provided for cosmic-ray removal (via
the `avsigclip' option in {\sc imcombine}) when combining them. The
final step was to subtract the co-added `off-object' frame from its
`on-object' counterpart, thereby (and quite trivially) producing an
image containing sky-subtracted spectra.

The exact location of each of these spectra on the image was determined
using the associated arc-lamp exposure, with a program written to
search for the pattern of 4 arc lines (8408.2, 8424.6, 8521.4 and
8667.9\AA) expected over our observed wavelength interval. Once
recognised, their centroids were used to define the bounds of each
spectrum in both the spatial and dispersion directions. Since the
1.5$''$ diameter `micro-slits' project to $\sim 5$ pixels on the
LDSS$++$ detector, the spatial bounds were set so as to include 2 CCD
rows either side of the row in which the mean centroid position was
located. In the dispersion direction, a calibration between wavelength
and CCD column was established using the centroids of the 4 arc lines;
this was used to define the bounds of each spectrum, being those
columns which lay within the interval $8365\leq \lambda\leq
8742$\,\AA\ (the limits of the blocking filter). With these bounds
defined, each spectrum was then extracted from the image and collapsed
down in the spatial direction to produce a final one-dimensional
sky-subtracted and wavelength calibrated spectrum. A total of 675
spectra were recovered in this process, providing observations of 586
unique galaxies (given the number of duplicate `double' slit spectra).

All of the spectra were then inspected independently by two of us
(M.L.B.\  \& W.J.C.), to search for H$\alpha$ emission. Any conspicuous
emission-line feature was checked in two ways: firstly, to see that it
was not residual night-sky emission that had not been fully removed in
the sky-subtraction process, and secondly, to confirm its
identification as H$\alpha$ by detection at the correct observed
wavelength of the nearby [N{\sc ii}]$\lambda\lambda$6548,
6583\AA\ lines (as also any other line; in particular
[SII]$\lambda\lambda 6716, 6731$\AA). The two separate lists of
identifications were then compared and this revealed 3 distinct
categories of detections: (i)\,clear cut cases where H$\alpha$ was
easily identified and confirmed by one or more other emission lines;
(ii)\,an emission line was clearly identified, but its confirmation as
H$\alpha$ via the presence of one or both of the [N{\sc ii}] lines was
highly dubious, (iii)\,an emission feature was seen but flagged as
being highly doubtful due it being either ambiguous (ie. there were
other equally significant but doubtful emission line features in the
spectrum) or a possible residual night-sky feature. In case (i) there
was no disagreement between the two identifiers. As might be expected,
there were more instances of disagreement for the case (ii) and (iii)
objects; these were resolved by further inspection and checking of
features and coming to a consensus on their identification. As a
result, there was some promotion and demotion of objects between the
two classes. The final merged list contained 43 case (i) objects and 80
case (ii) objects; any case (iii) object was at this point discarded
from any subsequent analysis. Representative examples of case (i) and
case (ii) spectra are shown in Fig.~3.

To quantify the emission observed from these objects, a flux
calibration was derived for our spectra using the same $I$-band image
upon which our target selection was based. It was used to calculate the
magnitudes objects would have when observed through our 1.5$''$
circular `micro-slits'\footnote{The fact that the $I$-band image and
our LDSS$++$ observations were taken in almost identical seeing
conditions ($\sim 1.3''$ FWHM) should ensure that this is done
accurately, free of any profile-dependent systematic effects.}. The
corresponding flux (in erg\,s$^{-1}$\,cm$^{-2}$\,\AA$^{-1}$) received
by the spectrograph was computed from these `slit' magnitudes using the
{\sc nicmos} converter, which in turn was used to determine the flux that
fell within the wavelength interval of each spectrum. This was compared
with the total number of continuum counts recorded in our observed
spectra (corrected for any scattered light; see below) to give a mean
flux (per count) calibration of $(8.7\pm
1.0)\times10^{-19}$\,erg\,s$^{-1}$\,cm$^{-2}$\,count$^{-1}$.  In
applying this to the data, the uncertainty of the flux in each pixel
was computed using the standard, Poisson treatment, including the read
noise and Poisson uncertainty resulting from the sky subtraction. We
also included the 12\% uncertainty in the flux calibration which was
added in quadrature.

One complication of the nod and shuffle technique is that scattered
light affects the object and sky spectra differently.  This introduces
a small systematic error in the continuum level of the extracted
spectra.  For a subset (88) of the spectra, the underlying scattered
light component was determined locally on the `subtracted' image by
taking the mean signal just above and below the extracted spectra and
averaging the two values. For these galaxies, a good correlation was
found between the mean (corrected) continuum flux near H$\alpha$ and
the total I band magnitude:  $\mbox{log}_{10}f_c=-8.03-0.46 I$, where
$f_c$ is the continuum flux in ergs s$^{-1}$cm$^{-2}$pix$^{-1}$ and the
rms scatter in the logarithm of this quantity was found to be 0.02. We
used this correlation to estimate the continuum level for the remaining
spectra, and the appropriate flux was added to the spectrum to achieve
this.

The H$\alpha$ flux was measured automatically for each spectrum, using
a {\sc fortran} program available upon request.  First, the continuum
was determined by fitting a single line to the pixels lying in the rest
wavelength ranges $6490<\lambda<6540$ and $6590<\lambda<6640$ (chosen
to avoid the [N{\sc ii}] lines), using weighted linear regression with
the flux in each pixel weighted by the inverse square of its
uncertainty.  An iterative rejection of points lying 1.5$\sigma$ away
from the fit was employed to exclude noise spikes and absorption
features. The H$\alpha$ flux was then computed by summing the flux
above this continuum, in the rest wavelength range $6556<\lambda<
6570$, which includes all the flux from our widest lines and stays well
clear of the adjacent [N{\sc ii}] lines.  For the pixels at the extreme
ends of this wavelength interval, we only include a fraction of the
flux, equal to the fraction of the pixel which lies within the
specified interval. The uncertainty in the line flux for each pixel is
then computed as the quadrature sum of the uncertainties in the total
flux and in the continuum fit at that pixel (including the correlation
coefficient in the latter).

We did not compute equivalent widths in the usual way, because the
continuum level is very low (consistent with zero) for many galaxies, and
because there are systematic uncertainties in its level due to the
scattered light problem discussed above. However, it is still of interest
to have a measure of a galaxy's fractional H$\alpha$ luminosity; that is,
how bright the H$\alpha$ line is compared with the total continuum flux of
the galaxy.  We therefore computed the ratio of the H$\alpha$ flux
to the I-band continuum flux.  In approximate terms, multiplying this
ratio by the rest-frame width of the H$\alpha$ line in our spectra
(typically $\sim 7$\AA) yields a value comparable to a rest-frame
equivalent width. 

The detection limit in this survey is a strong function of wavelength,
as the sky is much brighter at the blue and red ends of our spectral
range than it is in the centre.  In order to estimate a limit which
corresponds as closely as possible to the manner in which we detected
our lines, we repeatedly chose spectra at random from those in which we
did not detect an emission line. To each spectrum we added a gaussian
H$\alpha$ emission line at a random redshift (but within our filter
limits) and with a randomly selected flux.  The two [N{\sc ii}] lines
were also added with random fluxes, with the (fairly arbitrary)
additional restrictions that they each had less than 70\% of the flux
of the H$\alpha$ line, and had fluxes within 10\% of each other. This
whole process was done 1000 times. One of us (M.L.B.) then tried to
locate the H$\alpha$ line in each of these spectra, with an effort to
mimic the detection criteria for the real sample as closely as
possible.  We then determined the success rate in identifying these
lines as a function of line flux and redshift; in particular, we
identify the line flux at which our success rate was 50\%, in four
redshift bins.  For objects with $z>0.29$, the 50\% detection limit is
$\sim 4.3 \times 10^{-17}$\,ergs\,s$^{-1}$\,cm$^{-2}$; since the
detection rate rises steeply with flux for these galaxies, we adopt
this value as our flux limit. For galaxies with $z<0.29$, the 50\%
limit is about a factor of two brighter, due to the presence of several
bright sky lines.

\begin{figure*}
\centerline{\psfig{file=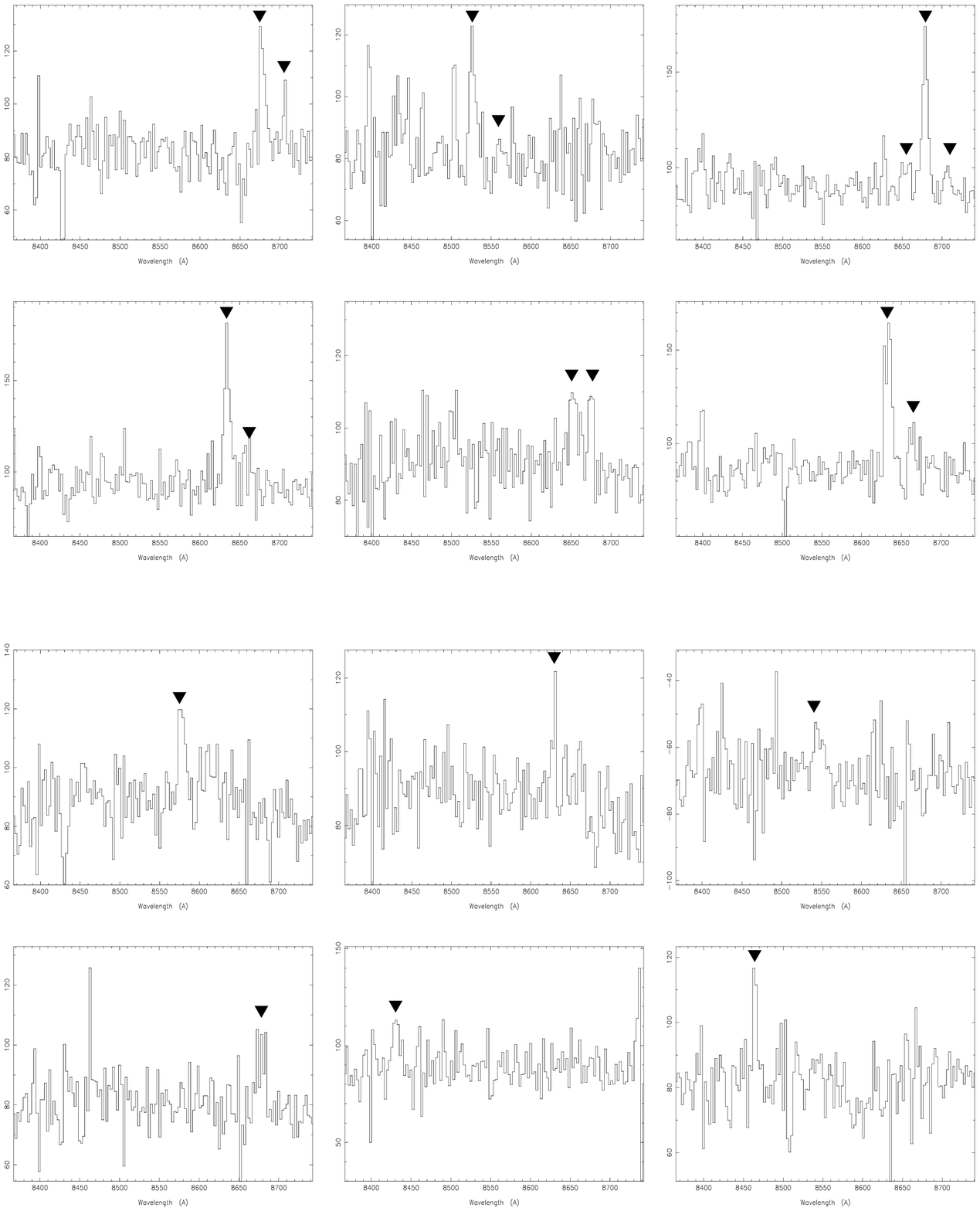,width=3.0in}}

\vspace{1.75in}

\caption{Representative spectra in which H$\alpha$ was detected. 
{\it Top two rows:} case (i) detections; the H$\alpha$ line and
its neighbouring (but weaker) [N{\sc ii}] line(s) are marked.
{\it Bottom two rows:} case (ii) detections; the H$\alpha$ line
is marked.} 
\end{figure*}

\section{Results}

\begin{figure*}
\centerline{\psfig{file=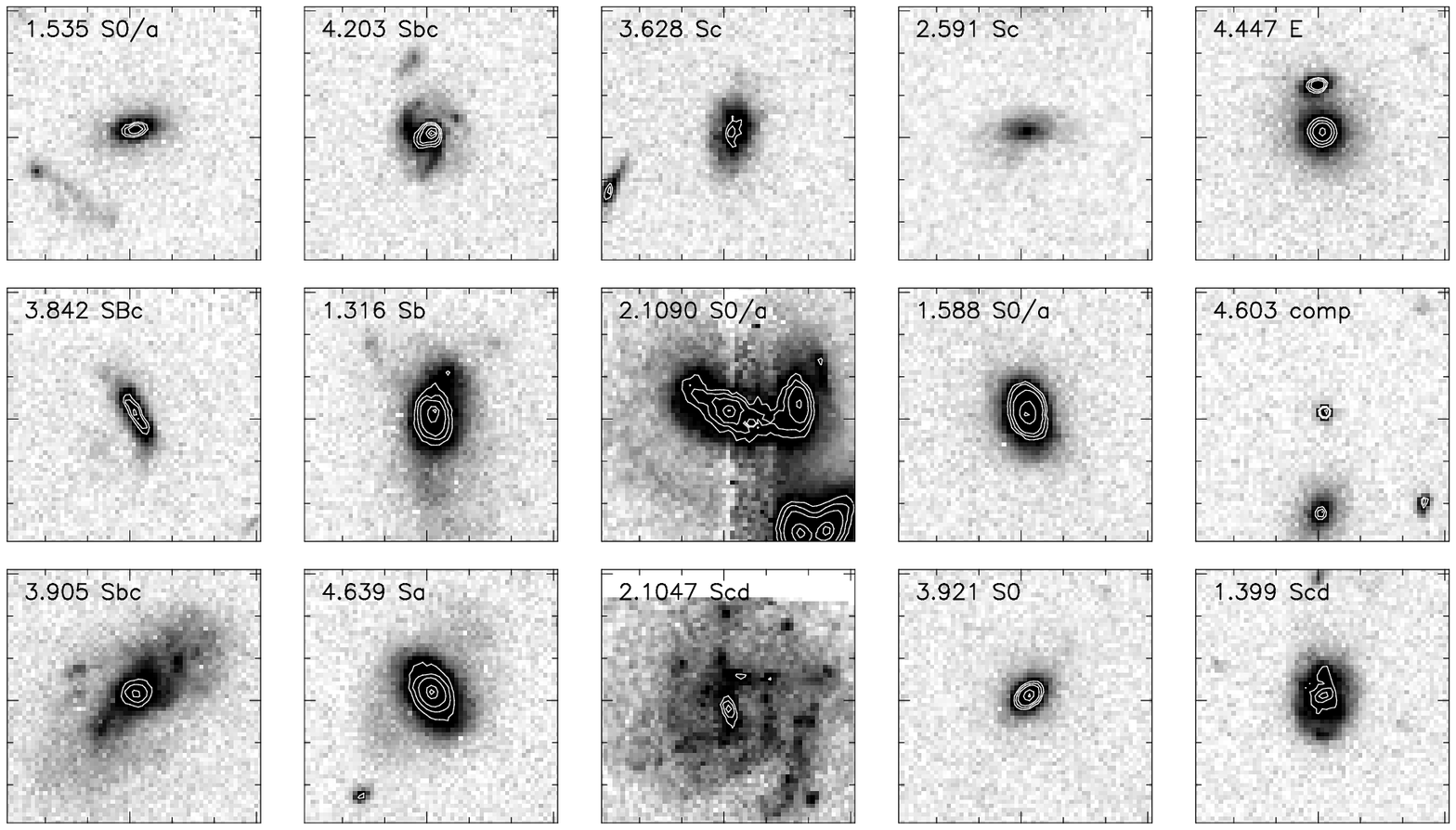,angle=270,width=6.5in}}
\caption{{\it HST} F702W images of a representative collection of the
H$\alpha$ emitters found in our survey. All the galaxies contained
in the top 2 rows have definite [case (i)] H$\alpha$ detections;
the galaxies in the bottom row have less certain [case(ii)] detections.
Each image is 6$''$ on a side; the {\it HST} identifications and
morphologies are shown at the top left.} 
\end{figure*}

A total of 51 cluster members -- 24 in the case(i) category and 27 in
the case (ii) category -- were found to be above our H$\alpha$ flux limit, 
and we now confine our attention to their properties. As a percentage
of the number of galaxies observed, these numbers are rather small;
at most the H$\alpha$ emitters make up only 9\% (51/586) of our sample, 
with only 4\% (24/586) being secure case (i) detections. Taking these
results for a single cluster at face value, the low detection rate
emphasises the need for spectroscopy with large multiplexing ($N\sim
10^{3}$) in order to obtain samples of these objects of modest size. 

The number of cluster members detected in H$\alpha$ and for which {\it
HST} morphologies are available is also small: 15 fall within the C98
WFPC2 mosaic and of these, 10 (67\%) were classified as spirals (1 Sa,
3 Sb, 6 Sc and later), 3 (20\%) as S0's or S0/a's, 1 as an elliptical,
and 1 a compact galaxy. The {\it HST} F702W images of a representative
collection of these objects is shown in Fig.~4.  While the majority are
spiral galaxies, as we might expect, it is of note that H$\alpha$
emission is not solely confined to these types with some of our
detections being in the supposedly dormant S0 and E galaxies.

The incidence of H$\alpha$ emission as a function of $I_{tot}$ magnitude
and radius from the centre of AC\,114 is shown in the top two panels of
Fig.~1. In both plots, the points connected by the solid lines show the
incidence of both case (i) and case (ii) types, while the dotted line
shows just the case (i) types. Note that in both cases the points have
been corrected for the varying sampling rate of `cluster' galaxies with
magnitude and radius, having been divided by the corresponding cluster
galaxy fraction as shown by the dotted line in the bottom two panels. In
constructing Fig.~1(d), we have used radial bins of unequal size to
ensure each point is based on approximately the same number of galaxies
and hence has equal statistical weight.

We see in Fig.~1(c) a marginally higher detection rate at brighter
magnitudes ($I_{tot}<20.5$) than at fainter magnitudes, and given the
way we have set our flux limit (see \S 3), this is likely to be real
and not due to any increase in incompleteness with magnitude. In
comparison, Fig.~1(d) is remarkable for how flat the H$\alpha$
detection rate is with radius, with no gradient seen at all from the
centre right out to $2\,h_{50}^{-1}$\,Mpc \footnote{Although we note
that there are {\it no} H$\alpha$--emitting galaxies within the central
250\,$h_{50}^{-1}$\,kpc -- something which is obscured by the binning
procedure used in Fig.~1(d)}. This can be contrasted with the
morphological content of AC\,114 observed by C98 over the central $\sim
1\,h_{50}^{-1}$\,Mpc. In particular, they found the fraction of {\it
spiral} galaxies to rise from 6 to $\sim 50$\% within the central
$0.5\,h_{50}^{-1}$\,Mpc and then remain approximately constant beyond
that; this translates to spiral fractions of 45\% and 50\% within the
two broad, innermost radial bins used in Fig.~1d (as shown by the 2
solid triangular points connected by the dot--dashed line). Clearly,
then, there are many spirals members of AC\,114 which have no significant
H$\alpha$ emission and thus ongoing star formation
(SFR$<0.25$M$_{\odot}$\,yr$^{-1}$); at radii of
$0.5-1.0\,h_{50}^{-1}$\,Mpc these make up of the order of two-thirds of
AC\,114's spiral population! This lends further support to the conclusion
of P99 that spiral-arm structure can persist in distant cluster spirals
for a considerable time after star formation has ceased
.

While we might expect the H$\alpha$ detection rate to increase with 
radius, it is plausible that any underlying cluster-centric variation  
simply gets washed out in Fig.~1(d) due to AC\,114's elongated
galaxy distribution. In Fig.~5(a), therefore, we plot the detection
rate as a function of local projected galaxy density\footnote{Based on
the circular area containing the 10 nearest neighbouring galaxies within
our SExtractor catalogue, with a correction applied for field galaxy
contamination.}, which will track galaxy `environment' much more closely
than radius. Once again we correct for the variation in the rate at which
`cluster' galaxies are sampled at different local densities. In this case
we do see evidence of a trend, with the detection rate at the lowest
densities being conspicuously (although, due to the large uncertainties,
not that significantly) higher, and the rate in the highest density bin
being marginally lower than that at intermediate densities. Hence we would
conclude that there is a link between the incidence of H$\alpha$ emission
and `environment' as measured in this way, although it is revealed only
rather weakly here.

As a further check, we plot the H$\alpha$ detection rate as a function
of galaxy color; here colors have been measured from a pair of $B$ and $R$
AAT prime-focus images also taken for another program. As we might
anticipate, we see a clear trend with the detection rate rising
monotonically with decreasing (bluer) color. This is to be expected given
that bluer galaxies are known to be more gas rich and active in star
formation, but it is of note that our detection rate does not drop to zero
for the reddest ($B-R\sim 2.5$) galaxies, with $\sim 5$\% of these
objects showing H$\alpha$ emission. These could well be the known
population of spiral galaxies whose colors are as red as the E/S0
population and which contribute to the red wing of the color distribution
(Butcher \& Oemler 1978b). Further insight into the distribution in 
color, as also luminosity, of the H$\alpha$-emitting population can be
seen in Fig.~6 where a color-magnitude diagram of the galaxies 
observed in this study is shown with the H$\alpha$ detections highlighted.
This emphasises further the tendency of H$\alpha$ emitters to be
amongst the bluer galaxies, but to also be present amongst the redder
galaxies, in particular those which populate the E/S0 color-magnitude
sequence (seen as the sloped ridge-line that runs across the diagram).
The figure also demonstrates that there is nothing peculiar
about their luminosity distribution with respect to the population
they were sampled from.  

In addition to the incidence of H$\alpha$ emission, its {\it strength} 
in galaxies at different locations within the cluster is also highly
pertinent to understanding the mechanism(s) by which star
formation is modulated in the cluster environment. The observed H$\alpha$
fluxes, $f$(H$\alpha$), have therefore been converted into 
luminosities and star formation rates (SFR) using the relations:
\[ L({\rm H}\alpha) = 
4.348\times 10^{57} h_{50}^{-2} f({\rm H}\alpha) [z(1+z/2)]^{2} ~ 
{\rm erg\,s}^{-1} \]
and 
\[ \rm{SFR} = \frac{{\sl L}({\rm H}\alpha)}{1.12\times 10^{41}} ~
{\sl E}({\rm
H}\alpha)~ ~ {\rm M}_{\odot}\,{\rm yr}^{-1}, \]
where the latter relation is taken from Kennicutt (1992). The quantity 
$E$(H$\alpha$) is a factor which corrects for the extinction suffered
at H$\alpha$; we have adopted here Kennicutt's canonical value of 2.5.
[The measured values of $f$(H$\alpha$), $L$(H$\alpha$), and SFR for 
those cluster members detected in H$\alpha$ are included in a full
listing of all the relevant data for the 586 unique galaxies observed
in this study, to be made available as a machine-readable table in
the on-line version of the paper. Other data included in this listing 
are: RA(J2000), Dec(J2000), $I_{tot}$, detection class and, where
available, Hubble (and T-) type, and $z$(H$\alpha$).]

The star formation rates calculated in this way are plotted as a
function of local galaxy density in Fig.~7. The extent to which the
star formation is suppressed in the cluster galaxies is striking with
the bulk of the objects in Fig.~7 populating the bottom of the diagram
(SFR's in the range 0.5--1.0\,M$_{\odot}$\,yr$^{-1}$), accompanied by a
sprinkling of objects above them but whose star formation rates do not
exceed SFR$\sim 4$\,M$_{\odot}$\,yr$^{-1}$. Furthermore, we see little
if any evidence for any gradient with projected galaxy density. It
would appear, therefore, that local galaxy density does not have any
detectable influence on cluster members' star formation rate.

\centerline{\psfig{file=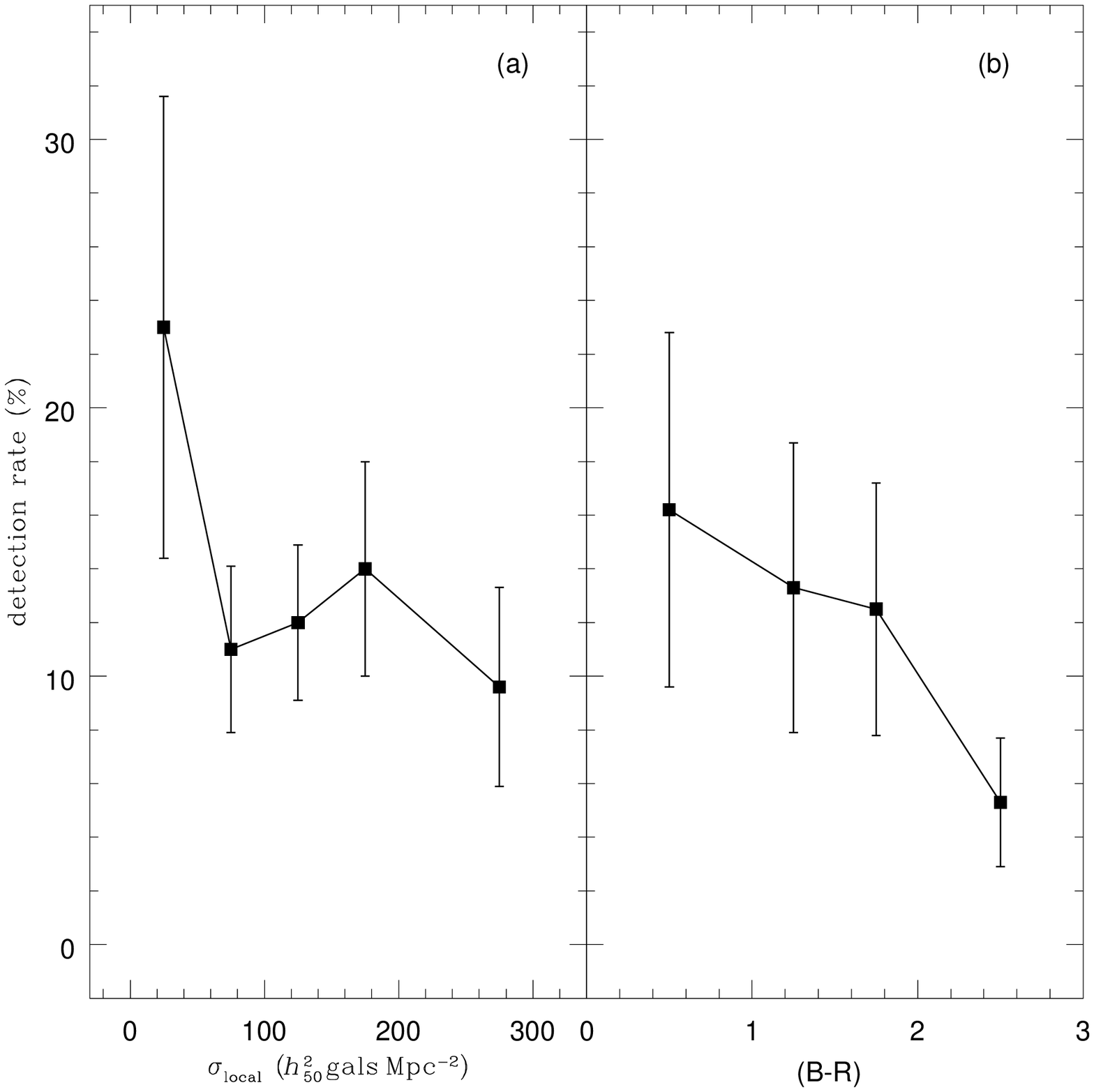,width=3.0in}}
\noindent{\small\addtolength{\baselineskip}{-1pt}  
{\sc Fig. 5.}-- The H$\alpha$ detection rate as a function of: (a) projected
local galaxy density, and (b) $(B-R)$ color.

} 

The upper bound to the star formation rates we see in AC\,114 of 
SFR$\sim 4$\,M$_{\odot}$\,yr$^{-1}$ -- which is about the same
level as that observed in the Milky Way (Rana 1991) -- contrasts quite
dramatically with the levels of star formation seen in field galaxies at 
the same redshift (Tresse \& Maddox 1998; hereafter TM), where many
galaxies have rates in the 5--10\,M$_{\odot}$\,yr$^{-1}$ range, with some
even higher than this. This can be seen more quantitatively in Fig.~8
where we compare AC\,114's H$\alpha$ luminosity function derived from this
work with that derived for field galaxies at $0.25\leq z\leq 0.35$ by
TM [based on the Canada-France Redshift Survey (CFRS) sample; Lilly et al.\
1996]. Here the `field' luminosity function has been normalised onto our
vertical scale via consideration of TM's detection rate over the same
limits in apparent magnitude ($I\leq 22.25$) and H$\alpha$ luminosity   
[$\log L($H$\alpha)>39$] that applied to this study. 
The difference between the two luminosity functions is very stark: 
AC\,114's function is {\it suppressed} by an order of magnitude with
respect to the field be it from the point of view of the numbers of
objects at a given H$\alpha$ luminosity or the H$\alpha$ luminosities at 
which the numbers of objects are equal. Moreover, AC\,114 is completely
deficient of galaxies with $L({\rm H}\alpha)>1.6\times
10^{41}$\,erg\,s$^{-1}$ and SFR$>4$\,M$_{\odot}$\,yr$^{-1}$.   

\section{Discussion}

The results of our analysis in the preceding section are quite
unambiguous: the fraction of galaxies with on-going star formation
in AC\,114 is uniformly low ($\sim 9$\% at most) right out to
$\sim 2\,h_{50}^{-1}$\,Mpc from its centre, and where star formation is
detected, the rates are highly suppressed with respect to those observed
in the `field' at the same redshift. 

None of the galaxies detected in this study have the high star
formation rates typically associated with `starburst' galaxies.  Of the
586 unique galaxies observed, we expect 380 to be cluster members; we
can therefore  determine that the fraction of starburst galaxies (with,
say, star formation rates greater than 10 $M_\odot$ yr$^{-1}$) in this
cluster is less than 0.3\% with 1$\sigma$ confidence (or 2.3\% at
3$\sigma$). This may seem surprising in light of the large number of
galaxies with apparently `post-starburst' spectra discovered in some
clusters (CS, C98 and Dressler et al.\ 1999).  However, this is not at
all a universal property, as there are clusters with very few
post-starburst galaxies (Balogh et al.\ 1999), and the fraction of such
galaxies in AC\,114 is not tightly constrained:  in combination, CS,
Couch et al.\ (1994) and C98 find $6/73=8\pm 3$\% of the observed
galaxies in AC\,114 have the blue colors, strong H$\delta$ lines and lack
of [O{\sc ii}] emission characteristic of post-starbursts.  It is
typically assumed that the lifetime of a starburst is about a factor of
ten smaller than that of the post-starburst phase, in which strong
Balmer lines persist for up to about 1 Gyr.   Our results would then
rule out a post-starburst fraction $\gs3$\% with 1$\sigma$ confidence
and $\gs23$\% with 3$\sigma$ confidence.  However, the bursts could
conceivably be much shorter, with lifetimes of only a few 100 Myr, in
which case the constraint becomes much weaker.  Alternatively, Balogh
\& Morris (2000) find that the fractions of post-starburst galaxies (as
determined from [O{\sc ii}] and H$\delta$ measurements) are
overestimated, because some of them show H$\alpha$ emission; none of
the three such galaxies in AC\,114 are detected in H$\alpha$ in the
present work, however.

\centerline{\psfig{file=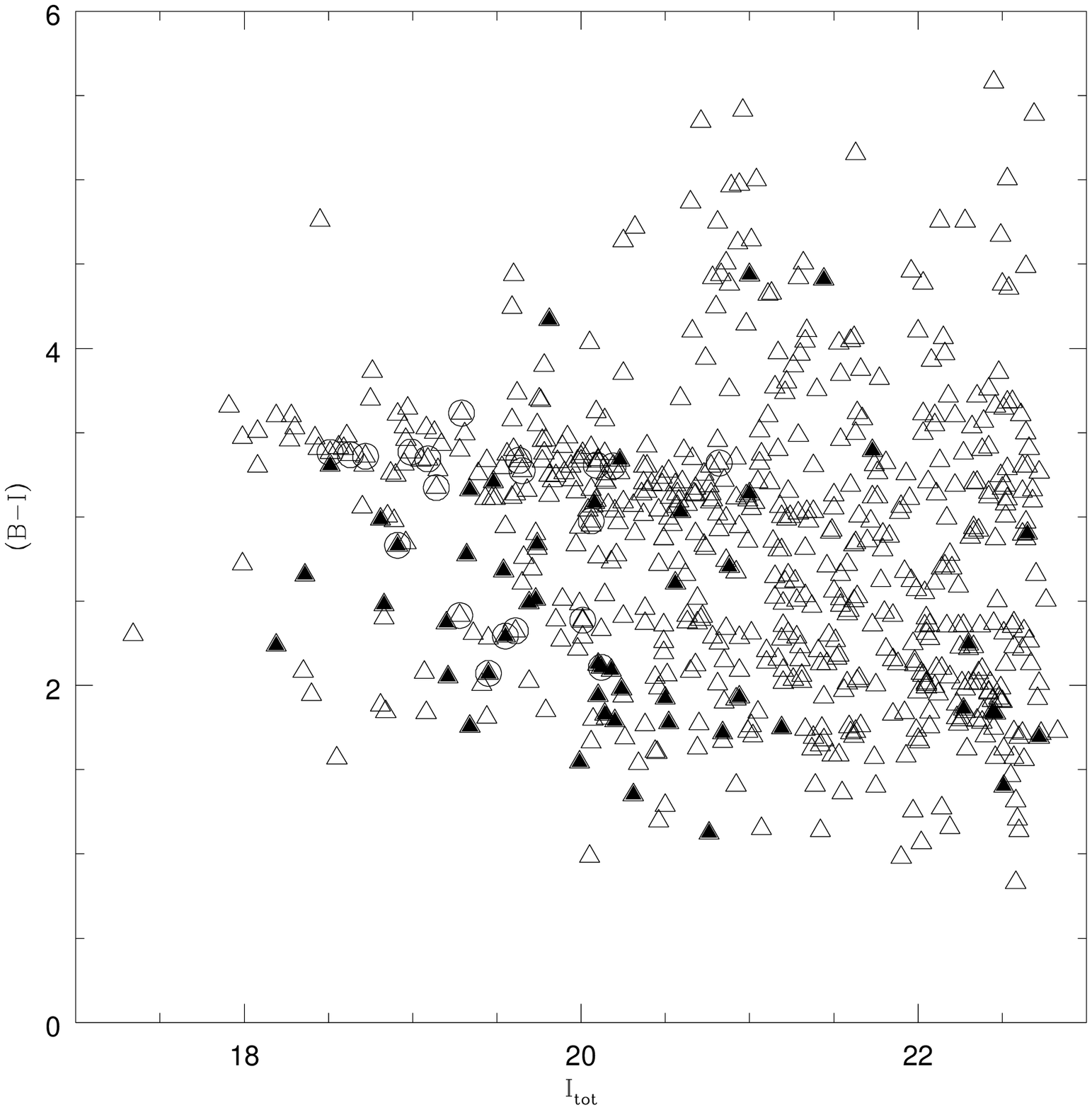,width=3.0in}}
\noindent{\small\addtolength{\baselineskip}{-1pt}  {\sc Fig. 6.}-- The distribution in the color-magnitude plane of our 
spectroscopic targets ({\it open triangles}); those plotted
as {\it filled triangles} indicate our H$\alpha$ detections.
Targets whose membership of AC\,114 had been confirmed by previous
spectroscopy are {\it circled}.

}

Both P99 and Smail et al.\ (1999) have cautioned that a considerable
amount of the star formation going on in distant cluster galaxies could
be obscured by dust.  Line equivalent widths will not be affected if
the dust is distributed as a uniform screen; however, they will be
reduced if the lines (produced in HII regions) are preferentially
extincted, relative to the continuum (P99).  Unfortunately, only three
of the galaxies detected in our sample have measurements of [O{\sc
ii}]$\lambda$3727 emission.  These three galaxies have EW([O{\sc
ii}])/EW(H$\alpha$ + [N{\sc ii}]) ratios of $0.4\pm 0.1$, $0.5\pm 0.2$
and $0.9\pm 0.3$; these ratios are not lower than the canonical value
of 0.4 observed for the `normal' nearby galaxy sample of Kennicutt
(1992), as would be expected if dust obscuration was important.  We
cannot rule out the possibility that there exists a population of
starburst galaxies in which the dust extinction is strong enough to
reduce the H$\alpha$ flux below our detection limit, though we note
that it would take about 4 magnitudes of extinction (at $\lambda=6563$
\AA) to obscure the H$\alpha$ flux from a galaxy with a star formation
rate of 10 $M_\odot$ yr$^{-1}$ in this way.

\centerline{\psfig{file=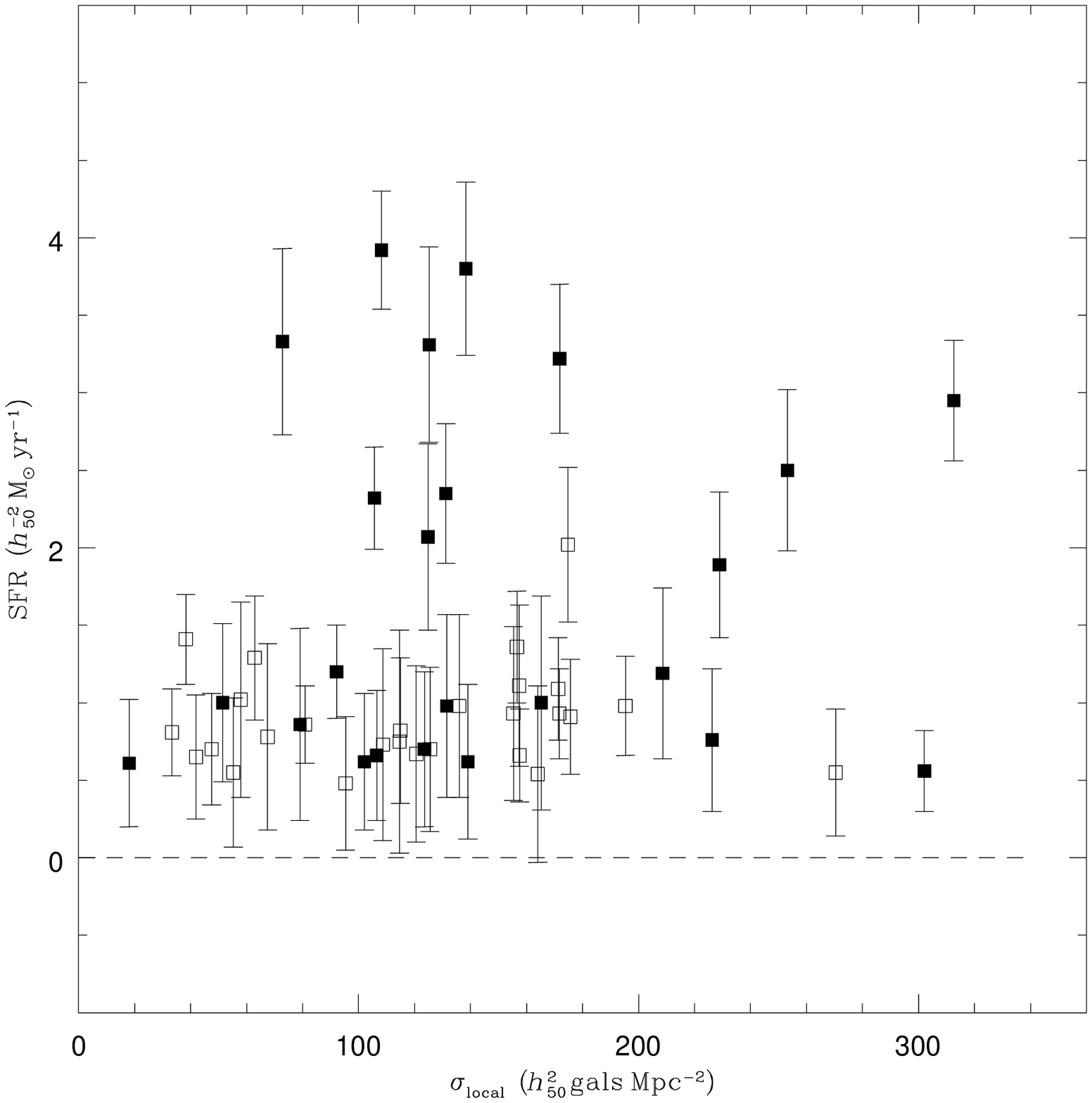,width=3.0in}}
\noindent{\small\addtolength{\baselineskip}{-1pt}  {\sc Fig. 7.}-- The star formation rate inferred from the observed H$\alpha$
flux plotted as a function of projected local galaxy density.
The {\it solid} squares indicate the secure, case(i) detections while
the {\it open} squares represent the less secure, case(ii) detections.

}

The general lack of `activity'  seen amongst galaxies observed in this
study is perplexing, as we are probing this cluster out to large
cluster-centric radii ($r\sim 2\,h_{50}^{-1}$\,Mpc), where the chances
of encountering recent arrivals to the cluster might be expected to be
higher. As the data of TM would indicate (see Fig.~6), such objects
should be conspicuous by their higher (on average) H$\alpha$
luminosities and thus star formation rates. The fact that we do not see
any sign of such a population suggests that either we have not gone out
far enough, or AC\,114 just happens to be a peculiar cluster in terms of
its structure (we have already noted its highly elongated galaxy
distribution), or perhaps both.

There has been recent theoretical work which suggests that the former
could certainly be the case. The modelling of star formation gradients
in distant clusters by Balogh et al.\ (2000) has shown that many of the
galaxies even at $2\,R_{vir}$\footnote{$R_{vir}$ is the virial radius
which we estimate to be $\sim 3\,h_{50}^{-1}$\,Mpc for AC\,114.} have
already passed through the cluster body one or more times, and
therefore cannot be considered recent arrivals.  A similar result is
found in the semi-analytic models of Diaferio et al.\ (2000), where
there is little gradient in the mean star formation rate of cluster
galaxies out to $R_{vir}$; beyond this radius it increases strongly,
but does not reach the field value until beyond 2$R_{vir}$.  Despite
the relatively wide field, the present observations still do not reach
the virial radius of AC\,114; it will be important to obtain observations
at even larger radii to map the star formation gradient further from
the cluster, until it matches the field properties, as this will
provide important constraints on models of cluster formation.

What is also very clear from the simulations is that the process of
cluster growth is a highly dynamic one with the episodic merging of
group and sub-cluster structures with the consequent mixing and 
`splashing' of galaxies. This serves to further concentrate galaxies
toward the centre of the cluster but at the same time fling a significant
number back out to larger radii. While the simulations allow this
behaviour to be followed with time and thus the time-averaged distribution
of infallers and the associated gradients in star formation to be
determined, the observations -- such as those of AC\,114 that we have
obtained here -- provide just a `snapshot' of the cluster assembly
process at these earlier epochs. While AC\,114 clearly offers a view of
a system where these dynamic processes have advanced sufficiently to
create a massive, high density, virialised core (Smail et al.\ 1997,
Natarajan et al.\ 1998), its pronounced elongation suggests its growth may
still be some way from being complete. Moreover, a more complete
observational picture will require many `snapshots' covering the entire
formation process; our observations of AC118 (a merging double mass
component system) and AC103 (a poor and irregular single component system)
which will be presented in forthcoming papers, will be important steps in
this direction.

\centerline{\psfig{file=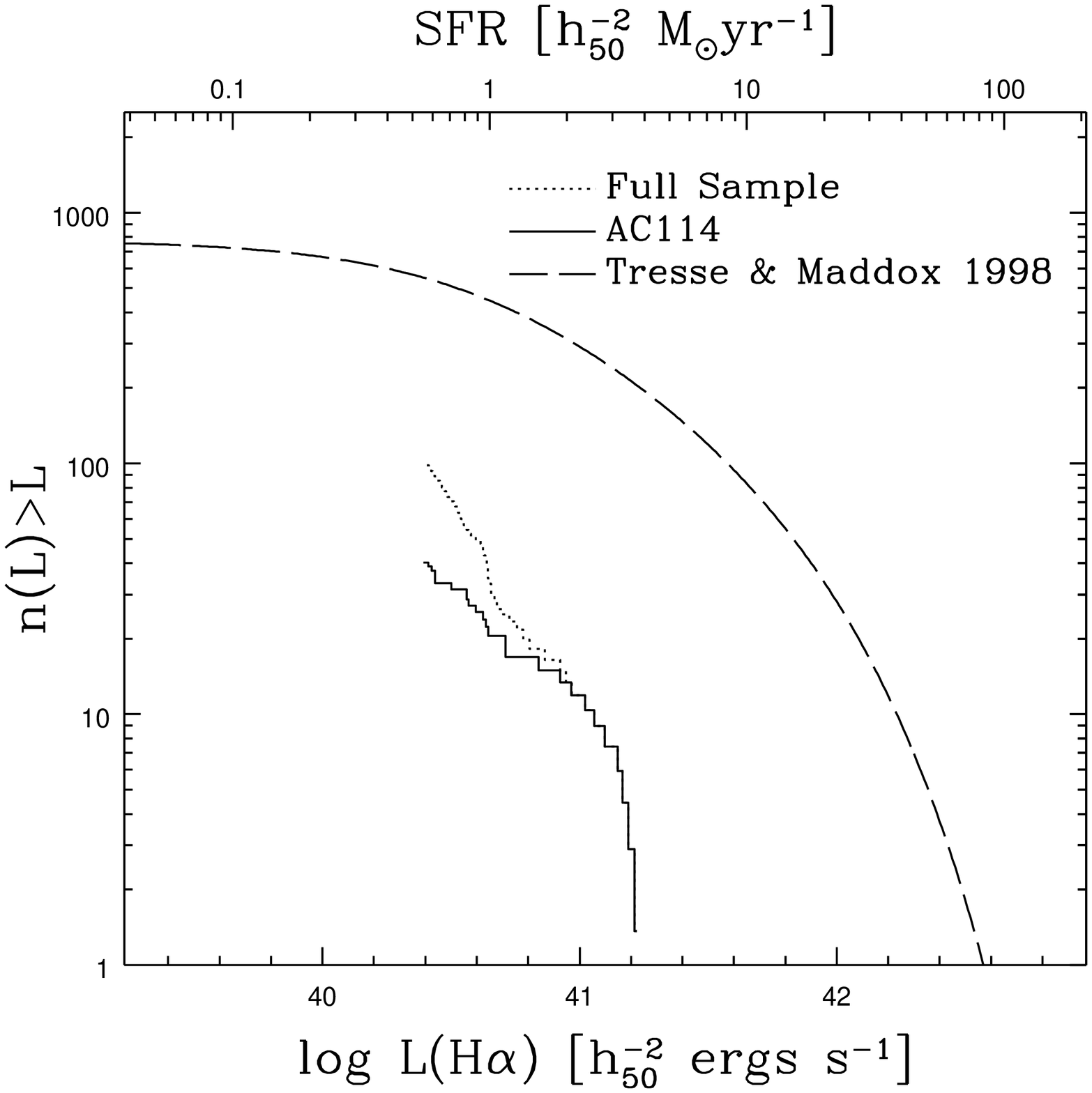,width=3.0in}}
\noindent{\small\addtolength{\baselineskip}{-1pt}  
{\sc Fig. 8.}-- A comparison of the H$\alpha$ luminosity function
measured here for AC\,114 with that of field galaxies at a similar
redshift measured from the CFRS sample by Tresse \& Maddox (1998). The
corresponding star formation rate is shown on the horizontal axis at
the top of the diagram.

} 

\section{Conclusions}

In this paper we have presented  H$\alpha$ spectroscopy for 586 unique
galaxies to $I\sim 22$ over a $3.1\times 3.1\,h_{50}^{-1}$\,Mpc field
centred on the rich galaxy cluster AC\,114 at $z=0.32$. The major
conclusions to be drawn from this survey of H$\alpha$ emitting galaxies
within this cluster can be summarised as follows:  \begin{itemize}
\item The incidence of H$\alpha$ emission in cluster galaxies is {\it
low} with 4\% of the galaxies surveyed here having highly secure
detections and the total being no more than 9\% if less secure
detections are included.  \item The galaxies in which H$\alpha$
emission is observed, have morphologies and colors that are largely
consistent with them hosting star formation activity. From the small
sub-sample of H$\alpha$ emitters which have {\it HST} morphologies, the
majority (67\%) are found to be spirals predominantly of later type.
H$\alpha$ emission is preferentially found in {\it bluer} galaxies
although we note that the detection rate does not drop to zero for the
reddest galaxies in the cluster.  \item In the central, $r\leq
1\,h_{50}^{-1}$\,Mpc region of AC\,114 where our {\it HST} imaging
overlaps with this study, approximately two-thirds of the galaxies with
spiral morphology were undetected in H$\alpha$. This provides further
evidence that morphological and star-formation evolution in clusters,
is largely decoupled.  \item Apart from the complete absence of
H$\alpha$--emitting galaxies within a radius of 250\,$h_{50}^{-1}$\,kpc
from the centre of AC\,114, the H$\alpha$ detection rate shows no
detectable radial variation out to $\sim 2\,h_{50}^{-1}\,$Mpc. However,
any radial trend will clearly be weakened by the highly elongated
structure of this cluster. A weak anti-correlation between detection
rate and {\it local} galaxy density is found, indicating that the
incidence of star--forming cluster galaxies increases by roughly a
factor of 2 in going from the highest ($\sim
250\,h_{50}^{2}$\,gals\,Mpc$^{-2}$) to the lowest ($\sim
50\,h_{50}^{2}$\,gals\,Mpc$^{-2}$) density regions probed within
AC\,114.  \item The rate of star formation inferred from the H$\alpha$
luminosities measured for the detected galaxies is found to be
uniformly low across the entire region of the cluster studied here. The
maximum SFR observed is $\sim 4$M$_{\odot}$\,yr$^{-1}$, with the
majority of galaxies having rates $< 1.5$M$_{\odot}$\,yr$^{-1}$. No
discernible variation in SFR with `environment' (as traced by local
galaxy density) was found.  \item A comparison of the H$\alpha$
luminosity function for galaxies in AC\,114 with those in the field at a
similar redshift shows the former to be highly suppressed. At the same
H$\alpha$ luminosity, AC\,114's luminosity function falls below that of
the field by {\it an order of magnitude}. Furthermore, galaxies with
luminosities $L({\rm H}\alpha)>10^{41}\,h_{50}^{-2}$\,erg\,s$^{-1}$
(and hence a SFR$>4$M$_{\odot}$\,yr$^{-1}$), are absent from the
cluster environment.

\end{itemize} 

\acknowledgements

This work was supported by the Australian Research Council. W.J.C.\
is grateful to the Physics Departments at Durham and St Andrews
Universities for their kind hospitality during the course of this
work. M.L.B.\ would like to thank the Department of Astrophysics at The
University of New South Wales for its hospitality during the completion of
this work, and gratefully acknowledges support from a PPARC rolling grant for
extragalactic astronomy and cosmology at the University of Durham.
I.R.S.\ acknowledges support through a Royal Society University Research
Fellowship.

\end{document}